\UseRawInputEncoding
\documentclass[]{aastex631}
\usepackage{amsmath}
\begin{document}

\title{Elemental Abundances at Coronal Hole Boundaries as a Means to Investigate Interchange Reconnection and the Solar Wind}

\correspondingauthor{Alexandros Koukras}
\email{alexandros.koukras@columbia.edu}

\author[0000-0002-6232-5527]{Alexandros Koukras}
% \email{alexandros.koukras@columbia.edu}

\author[0000-0002-1111-6610]{Daniel W. Savin}
% \email{savin@astro.columbia.edu}

\author[0000-0001-7748-4179]{Michael Hahn}
\affiliation{Columbia Astrophysics Laboratory, Columbia University, MC 5247, 550 West 120th Street, New York, NY 10027, USA}
% \email{mhahn@astro.columbia.edu}

\begin{abstract}

The origin of the slow solar wind is not well understood, unlike the fast solar wind which originates from coronal holes.
In-situ elemental abundances of the slow solar wind suggest that it originates from initially closed field lines that become open. 
Coronal hole boundary regions are a potential source of slow solar wind as there open field lines interact with the closed loops through interchange reconnection.
% Aim
Our primary aim is to quantify the role of interchange reconnection at the boundaries of coronal holes. 
% Methodology
To this end, we have measured the relative abundances of different elements at these boundaries. Reconnection is expected to modulate the relative abundances through the first ionization potential (FIP) effect. 
For our analysis we used spectroscopic data from the extreme ultraviolet imaging spectrometer (EIS) on board Hinode. To account for the temperature structure of the observed region we computed the differential emission measure (DEM). Using the DEM we were able to infer the ratio between coronal and photospheric abundances, known as the FIP bias.
% Results
By examining the variation of the FIP bias moving from the coronal hole to the quiet Sun, we have been able to constrain models of interchange reconnection. The FIP bias variation in the boundary region around the coronal hole has an approximate width of 30-50 Mm, comparable to the size of supergranules.
This boundary region is also a source of open flux into interplanetary space. 
We find that there is an additional $\sim$ 30$\%$ open flux that originates from this boundary region. 

\end{abstract}

\keywords{Solar physics(1476) --- Solar coronal holes(1484) --- Solar wind(1534)}

\section{Introduction} \label{sec:introduction}

The solar wind, the flow of charge particles emanating from the Sun, can be broadly categorized based on its speed as fast and slow solar wind \citep{Schwenn1990,Geiss1995,Zurbuchen2007}. Although it has been studied extensively, a complete, self-consistent, description has yet to be achieved \citep{Viall2020}.
To fully understand the solar wind it is necessary to determine the source regions for the different types of wind and the physical processes that accelerate them. 
The source of the fast solar wind has long been identified to be regions of open magnetic field, which appear dark in extreme ultraviolet (EUV) emission and are called coronal holes \citep[CHs;][]{Krieger1973,Zirker1977}.
In contrast, there is not a clear consensus for the source region of the slow solar wind \citep{Abbo2016}. 

In-situ observations, particularly those related to elemental abundances, indicate that the slow solar wind is composed of material that was initially confined within closed magnetic field lines and subsequently released onto open filed lines \citep{Gloeckler1989a, Geiss1995}.
This process can take place at the interface between open and close magnetic field concentrations. A potential source of the slow solar wind is the coronal hole boundary (CHB) regions. At this interface, closed magnetic loops, characteristic of the quiet Sun (QS) corona can reconnect with the open magnetic field lines of the coronal hole. This reconnection, typically referred to as interchange reconnection, allows plasma previously trapped in closed loops to be released into the solar wind. 
There are a number of theories \citep{Fisk2001,Antiochos2011} as well as observational evidence \citep{Zhao2009} that support this process.

\begin{figure}
    \centering
    \includegraphics[scale=0.5]{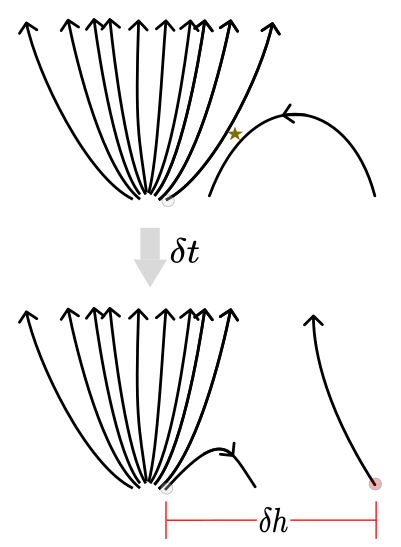}
    \caption{Graphical representation of the interchange reconnection that takes place at the boundary regions of coronal holes. The characteristic distance that the footpoint location of an open field lines shifts after reconnection is shown as the distance $\delta h$, and $\delta t$ represents the characteristic time for this process.}
    \label{fig:CHB_cartoon}
\end{figure}

One of these theories is the open-field diffusion model \citep{Fisk2001,Fisk2005,Fisk2006}, which indicates that coronal hole boundary regions are a source region of the slow solar wind.
The authors of this model argue that the motion of the open flux on the Sun is mainly governed by a diffusive process. Convective motions, such as differential rotation, have a smaller contribution in the movement of open flux and depend on the phase of the solar cycle and the position on the Sun.
The diffusive process is facilitated by the reconnection of open field lines with randomly oriented loops. This reconnection causes the creation of new loops and changes in the footpoints of the open field lines. 
Moreover, the diffusion model provides a way to quantify the diffusion of the open flux in the corona by establishing a diffusion equation \citep[see Equation\ 8 of ][]{Fisk2001} and a diffusion coefficient
\begin{equation}
    \kappa = \frac{(\delta h)^2}{2 \delta t} , \label{eq:diff_coeff}
\end{equation}
where $\delta h$ is the characteristic distance by which the footpoint of an open field line changes after reconnection with a close loop and $\delta t$ is the characteristic time for this process.
A graphical example of this process is shown in Figure~\ref{fig:CHB_cartoon}.

The measurement of elemental abundances at coronal hole boundary regions provide an observational signature to constrain the diffusion coefficient and study interchange reconnection. This diagnostic stems from the fact that closed field lines build up enhanced abundances of elements with a low first ionization potential (FIP), i.e., below 10~eV. This is known as the FIP effect \citep{Feldman2002,Feldman2007} and is quantified by the FIP bias. This is the ratio of the coronal-to-photospheric abundances for element $X$ and is defined as
\begin{equation}
   f_X = \frac{A_X^{\rm C}}{A_X^{\rm P}}, \label{eq:fip_bias}
\end{equation}
where $A_X^{\rm C}$ and $A_X^{\rm P}$ are the coronal and photospheric abundances of element $X$, respectively. 

Although there is still some debate about the exact mechanism that generates the FIP effect, the most widely accepted theory is that Alfv\'en waves, which can reflect or refract in the chromosphere, exert an upward pondermotive force that acts only on the ions \citep{Laming2004,Laming2009}. 
In the chromosphere, these ions are formed via photoionization by hydrogen Lyman-$\alpha$ radiation ($\approx 10.19$~eV) coming from the underlying photosphere \citep{vonSteiger1989}, explaining the 
observed 10~eV threshold for low- and high-FIP elements. This selective ionization results in low-FIP elements being more affected by pondermotive forces than high-FIP elements, generating the FIP effect. 

The FIP effect in the solar corona has been observed for many years. It displays consistent and well-established properties. The main ones are that the FIP effect occurs only in plasma confined by closed field lines and that the effect builds up linearly with the time that a field line remains closed. This time dependence has been observed on active region loops, where the FIP bias of low-FIP elements has been found to increase at a rate of 1/day \citep{Widing2001}. This property causes different solar features to have different FIP-bias values, based on the lifetimes of the loops that constitute them. The typical values for active regions are $>5$, for quiet Sun $\approx 4$ and coronal holes around $1-2$.

We expect, therefore, that there must be a transition in the FIP-bias values when we cross from one solar feature to another, as is the case at the boundaries between coronal holes and quiet Sun. This transition in FIP-bias values reflects also a transition in the magnetic field lines from open (coronal hole) to closed (quiet Sun). Consequently, by examining the characteristics of the transition in the FIP-bias values we can study the interchange reconnection that takes place at this boundary and try to constrain quantitative features of interchange reconnection models, such as the diffusion coefficient.

The opening and closing of magnetic fields lines in a boundary region around the coronal holes affects also the amount of open flux in this region. The quantity of this open flux has significant implications for what is known as the missing open flux problem. Essentially, the amount of solar open flux measured in situ, in interplanetary space, is significantly higher than the amount estimated from remote sensing observations and modeling. The main suspects for reconciling this issue have been the accuracy of the photospheric magnetograms and the precision of the open flux area estimates on the Sun. 
In this work, we address another potential source of open flux that has been little considered and could help to resolve the discrepancy, namely the boundary regions of coronal holes. By examining the interchange reconnection at the boundary region of a coronal hole and calculating the width of this boundary region, we are able to derived a first order estimate for the amount of open flux in this region.  

Here, we report our study of interchange reconnection at the boundary regions of coronal holes. Using spectroscopic observations we determine the FIP bias in the area surrounding the coronal hole and derive the width of the boundary where FIP-bias values show a significant change. 
The observations that we used are presented in Section~\ref{sec:observations} and the data processing we applied is discussed in Section~\ref{sec:data_analysis}. Our methodology for the derivation of the FIP bias is described in Section~\ref{sec:fip_bias} and our results are presented in Section~\ref{sec:results}. 
In Section~\ref{sec:discussion} we discuss the implications of this analysis and 
in Section~\ref{sec:conclusions} we present our concluding remarks.

\section{Observations} \label{sec:observations}

\begin{figure}[ht]
    \centering
    \includegraphics[scale=0.4]{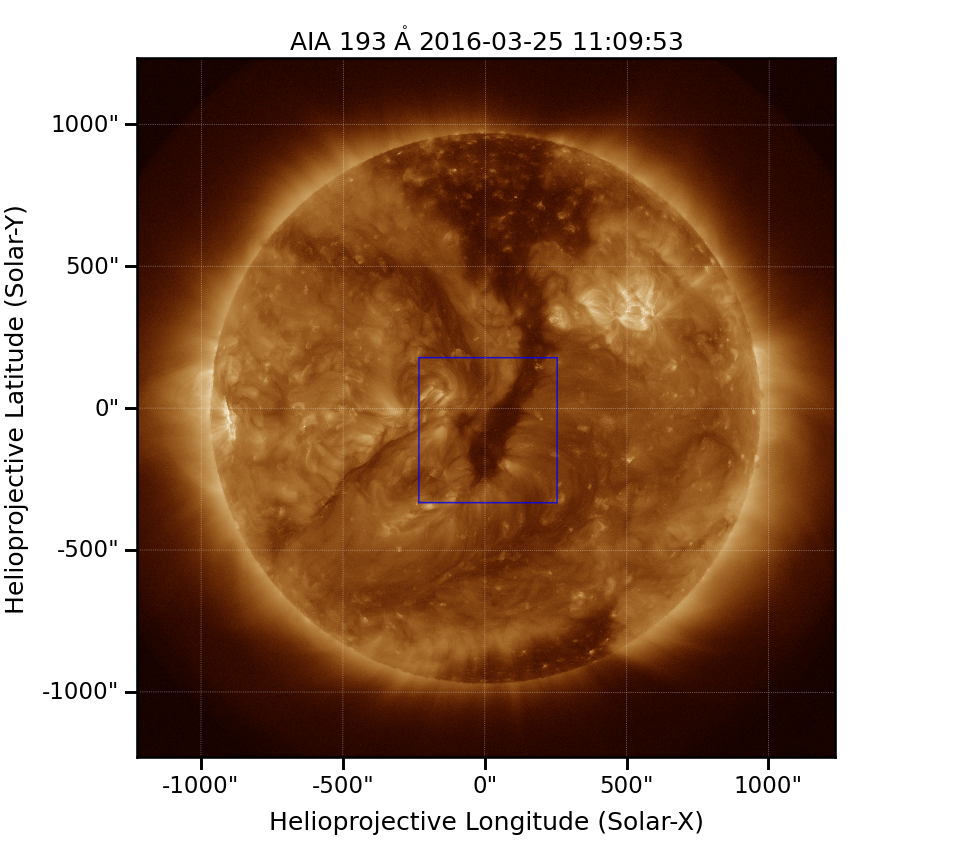} 
    \caption{The FOV of EIS, overlaid on the full disk image of the Sun. The background solar image is from the AIA/SDO 193~\AA\ observation closest to the midpoint in time of the EIS raster and consists mainly of emission from Fe \textsc{xii} and \textsc{xxiv}. The blue box outlines the raster range covered by the EIS 2\arcsec~slit observation on 2016 March 25 at 10:06 UT.}
    \label{fig:EIS_FOV_AIA}
\end{figure}

The observations used here were taken with the extreme ultraviolet imaging spectrometer \citep[EIS;][]{Culhane2007} on board Hinode \citep{Kosugi2007}.
The data were collected with the 2$\arcsec$ slit. The observations were performed on 2016 March 25, starting at 10:06 UT and 
ending at 12:12 UT. 
This was during the declining phase of solar cycle 24.
The slit was rastered from $X = -217\arcsec.18$ to $X = 250\arcsec.90$ while centered vertically at $Y = -66\arcsec$. The step of the slit during the scan was $4\arcsec$, creating 123 pointings, which result in a field of view (FOV) in longitude $\mathrm{FOV}_X = 512\arcsec$ and in latitude $\mathrm{FOV}_Y = 492\arcsec$, centered at $X = 9\arcsec$ and at $Y = -66\arcsec$, respectively. The exposure time for each pointing was 60~s.
The FOV was centered around an equatorial coronal hole but also covers part of the quiet Sun on either side of the coronal hole. Figure \ref{fig:EIS_FOV_AIA} displays the FOV of EIS, on top of the closest-in-time 193~\AA\ image from the Atmospheric Imaging Assembly \citep[AIA;][]{Lemen2012} on board the Solar Dynamics Observatory \citep[SDO;][]{Pesnell2012}. 

The EIS observation had 11 spectral windows. From these windows, we selected multiple lines of low-FIP elements such as iron and silicon and one high-FIP element line, that of sulfur. 
It must be noted here that the FIP of sulfur is 10.3~eV, making it more of an intermediate FIP line, but due to lack of additional high-FIP lines we consider it as our high-FIP line for the derivation of the FIP bias (see Section~\ref{sec:fip_bias}). 
This assumption is supported by the fact that sulfur acts as a high-FIP element in closed loops \citep{Laming2019} and is commonly used for spectroscopic studies of the FIP bias \citep{Guennou2015,Baker2018}. 
The list of all the lines that we used is given in Table \ref{tab:lines_list}. 

\begin{table}[]
    \centering
    \caption{List of Spectral Lines Used}
    \begin{tabular}{ccc}
    
        \hline \hline
        Ions & Wavelength (\AA) & log(T[K])\\
        \hline 
        Si \textsc{x}       & 258.375 & 6.15\\
        \hline 
         Fe \textsc{viii}   & 185.213 & 5.80\\
         Fe \textsc{ix}     & 188.497 & 5.95\\
         Fe \textsc{x}      & 184.536 & 6.05\\
         Fe \textsc{xi}     & 188.216 & 6.10\\
         Fe \textsc{xii}    & 195.119 & 6.20\\
         Fe \textsc{xiii}   & 202.044 & 6.25\\
         Fe \textsc{xv}     & 284.160 & 6.35\\
         \hline 
         S  \textsc{x}      & 264.233 & 6.20\\
        \hline 
    \end{tabular}
    \tablecomments{The emission lines that were used in our analysis are listed by elements and are sorted 
    based on the peak temperature of their contribution function.}    
    \label{tab:lines_list}
\end{table}

\section{Data analysis} \label{sec:data_analysis}

    \subsection{Line Fitting} \label{subsec:line_fitting}
    In order to derive the intensities of the selected lines from the EIS observation, we fitted Gaussian profiles to the spectral data for each line. We used the
    eispac\footnote{\url{https://eispac.readthedocs.io/en/latest/}} Python package \citep{Weberg2023}, created by the EIS team, which uses the Minpack-1 FIT (MPFIT) method to solves the non-linear least squares problem with the Levenberg-Marquardt algorithm. 
    
    Most of the selected lines are unblended; in which case a single Gaussian fit was applied to the data. 
    There were a few blended lines in our data set, but they could readily be analyzed using multi-Gaussian fits.
    The Fe \textsc{xii} 195.12~\AA\ line is self-blended with the Fe \textsc{xii} 192.18~\AA\ line. A double Gaussian fit, as described by \citet{Young2009}, is sufficient to extract the $\lambda195.12$ line. 
    The Fe \textsc{xi} 188.216~\AA\ is self-blended with the Fe \textsc{xi} 188.299~\AA\ line. Similarly, using a double Gaussian fit we can derive the Fe \textsc{xi} 188.216~\AA\ intensity.
    The Fe \textsc{ix} 188.497~\AA\ line is blended with the S \textsc{xi} 188.675~\AA\ line and the Ar \textsc{xi} 188.806~\AA\ line. For this case we used 3 Gaussians to fit the spectral profile and derive the Fe \textsc{ix} intensity. 

    \subsection{Density Diagnostics} \label{subsec:density_diagn}
    In order to calculate the FIP bias, which is discussed in more detail in Section~\ref{sec:fip_bias}, it is important to estimate the electron density inside our FOV. 
    This can be done spectroscopically using density sensitive ratios of emission lines from the same ion \citep{Kenneth2008}. 
    The inferred density is determined using the emissivity ratio plots from databases of atomic data such as CHIANTI \citep{Dere1997}.     
    
    From the available lines in our dataset, the only suitable density diagnostic was the Fe \textsc{xiii} 203.82~\AA/202.04~\AA~ratio. The $\lambda203.82$ line is a self-blend of the $\lambda203.797$ and $\lambda203.828$ Fe \textsc{xiii} lines. The identification of these lines sometimes can be hindered by a blend with the Fe \textsc{xii} line at 203.72~\AA; but typically a double Gaussian fit is sufficient to separate the Fe \textsc{xiii} and Fe \textsc{xii} components.
    Next, considering that $\lambda203.797$ and $\lambda203.828$ have a ratio of 0.366, determined by the underlying atomic physics \citep{Dere1997}, we can derive the intensity of the $\lambda203.82$ line \citep{Young2007}. 
    The emissitivity ratio plot was calculated using CHIANTI version 10.0 \citep{DelZanna2021} and an average density of $\sim$ $10^{8.5}~\mathrm{cm}^{-3}$ was found for the whole FOV. Since the computed density varied little across the different areas in the FOV, and especially for the QS regions, we decided to use the average density for all pixels.

\section{FIP bias measurements} \label{sec:fip_bias}

Spectroscopic measurements of the FIP effect rely on the fact that the radiance (typically referred to as intensity) of a spectral line is proportional to the abundance of the element forming the line.\footnote{The radiance of a line is linearly proportional to the column density} 
The intensity $I$ of an optically thin spectral line, emitted by the transition of an electron from a higher energy level $j$ to a lower energy level $i$ is given by \citet{Aschwanden2005_book} as
\begin{equation}
\label{eq:Iji}
    I_{ji} = A_X  \int C_{ji}(T_{\rm e},n_{\rm e}) n_{\rm e}^2 dz,
\end{equation}
where $A_X = n(X)/n(H)$ is the abundance of element $X$ relative to hydrogen, $C_{ji}$ is the so-called contribution function, $T_{\rm e}$ is the electron temperature, $n_{\rm e}$ is the electron density, and the integration in $z$ is along the line of sight (LOS).

The contribution function\footnote{Some authors define the contribution function as $G_{ji}(T_e,n_e) = A_X C_{ji}(T_e,n_e)$} is defined as
\begin{equation}
    C_{ji}(T_e,n_e) = \frac{h\nu_{ji}}{4\pi} \frac{A_{ji}}{n_e} \frac{n_j(X^{+m})}{n(X^{+m})} \frac{n(X^{+m})}{n(X)} \frac{n(H)}{n_e}. \label{eq:C_ij}
\end{equation}
Here, $h$ is the Planck constant; $\nu_{ji}$ is the frequency of the emitted radiation; $A_{ji}$ is the Einstein coefficient for the spontaneous emission probability for $j \to i$; $n_j(X^{+m})$ is the number density of level $j$ of the $+m$ charge state of element $X$; $n(X^{+m})$ is the number density of the ion, and $n(X)$ is the total number density of element $X$.

For most spectral lines, the contribution function has a strong dependence on temperature and only a weak dependence on density, which enables one to treat the density as constant \citep[][]{Pottasch1963,Pottasch1964}. One can then re-express Equation~(\ref{eq:Iji}) as an integral over a temperature range with the form 
\begin{equation}
    I_{ji} = A_X  \int C_{ji}(T_{\rm e}) {\rm DEM}(T_{\rm e}) dT_{\rm e}. \label{eq:line_intensity_final}
\end{equation}
Here ${\rm DEM}(T_{\rm e}) = n_{\rm e}^2 dz/dT_{\rm e}$ is the differential emission measure and provides a measure of the amount of the emitting plasma along the LOS as a function of temperature \citep{Craig1976}. 

In order to calculate the DEM for a set of observed spectral lines it is necessary to invert Equation~(\ref{eq:line_intensity_final}). 
This is, by nature, an ill-posed problem.\footnote{Does not satisfy the three Hadamard criteria: (1) have a solution, (2) have a unique solution, and (3) have a continuous (stable) solution.} A number of techniques have been developed to address this problem, which are based on integral inversion methods \citep{Goryaev2010,Hannah2012,Hannah2013,Cheung2015,Plowman2020,Massa2023}.
For our analysis and for the results that are presented in Section~\ref{sec:results} we used the regularization method from \citet{Hannah2012,Hannah2013} to compute the DEM. The regularization method limits the amplification of uncertainties by introducing a ``smoothness'' criterion, allowing a stable inversion and a quick recovery of the DEM solution. 

    \subsection{FIP bias measurement using the DEM} \label{subsec:fip_bias_dem}
    For our DEM inversion, we use a set of spectral lines that are sensitive to the FIP effect. Combining the resulting DEM with a line that is insensitive to the FIP effect provides a useful method for determining the FIB bias \citep{Baker2013,Guennou2015}. 
    
    Following \citet{Guennou2015}, the first step is to compute the DEM only using low-FIP lines (typically only iron lines).
    The observed line intensity is given by 
    \begin{equation}
        I_{X_{\rm{L\!F}}}^{\rm{obs}} = A^{\rm{C}}_{X_{\rm{L\!F}}} \int C_{X_{\rm L\!F}}(T_{\rm e}){\rm DEM^C}(T_{\rm e}) dT_{\rm e}, \label{eq:Iobs_DEMc}
    \end{equation}
    where ${\rm DEM^C}$ is the coronal DEM.
    However, since the coronal abundances of the low-FIP lines are unknown a priori, it is useful to re-express the observed intensities based on the known photospheric abundances \citep{Grevesse1998}. 
    Using Equation~(\ref{eq:fip_bias}),
    we can rewrite Equation~(\ref{eq:Iobs_DEMc}) as 
    \begin{equation}
        I_{X_{\rm{L\!F}}}^{\rm{obs}} = A^{\rm{P}}_{X_{\rm{L\!F}}} \int C_{X_{\rm L\!F}}(T_{\rm e})f_{X_{\rm{L\!F}}}{\rm DEM^C}(T_{\rm e}) dT_{\rm e}. \label{eq:Iobs_DEMcp}
    \end{equation}
    We can further transform this equation 
    using the photospheric abundances DEM$^{\rm P}$, which is given as
    \begin{equation}
        {\rm DEM^P}(T_{\rm e}) = f_{X_{\rm{L\!F}}} {\rm DEM^C}(T_{\rm e}). \label{eq:DEMs}
    \end{equation}
    Substituting Equation~(\ref{eq:DEMs}) into Equation~(\ref{eq:Iobs_DEMcp}) yields
    \begin{equation}
       I_{X_{\rm{L\!F}}}^{\rm{obs}} = A^{\rm{P}}_{X_{\rm{L\!F}}} \int C_{X_{\rm L\!F}}(T_{\rm e}){\rm DEM^P}(T_{\rm e}) dT_{\rm e}. \label{eq:Iobs_DEMp}
    \end{equation}
    We can then invert Equation~(\ref{eq:Iobs_DEMp}) to determine ${\rm DEM^P}$.  Next, using the inferred DEM$^{\rm P}$, enables us to simulate the intensity of any specific line
    \begin{equation}
        I_{X}^{\rm{sim}} = A_{X}^{\rm{P}} \int C_{X}(T_{\rm e}){\rm DEM^P}(T_{\rm e}) dT_{\rm e}.
    \end{equation}
    By comparing the simulated and observed intensities of a spectral line from an element with a FIP that is different from the low-FIP element, we can calculate the relative FIP bias for any set of elements. 
    For the case of high- and low-FIP lines, the FIP bias ratio can be calculated using
    \begin{equation} \label{eq:I_sim/I_obs}
        \begin{split}
            \frac{I^{\rm{sim}}_{X_{\rm{H\!F}}}}{I^{\rm{obs}}_{X_{H\!F}}} & = \frac{A_{X_{\rm{H\!F}}}^{\rm{P}} \int C_{X_{\rm{H\!F}}}\rm{DEM^P} dT_{\rm e}}{A_{X_{\rm{H\!F}}}^{\rm{C}} \int C_{X_{\rm{H\!F}}}\rm{DEM^C} dT_{\rm e}} \\
            & = \frac{A_{X_{\rm{H\!F}}}^{\rm{P}} f_{X_{\rm{L\!F}}}\int C_{X_{\rm{H\!F}}}\rm{DEM^C} dT_{\rm e}} {A_{X_{\rm{H\!F}}}^{\rm{C}} \int C_{X_{\rm{H\!F}}}\rm{DEM^C} dT_{\rm e}} \\
            & = \frac{f_{X_{\rm{L\!F}}}} {f_{X_{\rm{H\!F}}}}.
        \end{split}
    \end{equation}

    In order to accurately constrain the DEM it is necessary to have a selection of lines formed over a large range of temperatures. 
    To improve the accuracy of the DEM in the temperature range where our high-FIP line is formed, we explored adding a line from the low-FIP element Si, which has a FIP similar to ${\rm{Fe}}$, 8.2~eV and 7.9~eV respectively.
    After calculating $\rm{DEM}^{\rm{P}}$ using only the available Fe lines, we compared the simulated and observed intensities of the Si \textsc{x} 285.375 \AA\ line
    \begin{equation}
        \frac{I^{\rm{sim}}_{\rm{Si}}}{I^{\rm{obs}}_{\rm{Si}}} = \frac{f_{\rm{Fe}}}{f_{\rm{Si}}}. \label{eq:Fe_Si_fipbias}
    \end{equation}
    For Equation~(\ref{eq:Fe_Si_fipbias}) we found an average value of $1.05 \pm 0.76$ inside our FOV. This indicates that Fe and Si can be assumed to have the same FIP bias ($f_{\rm{Fe+ Si}} \equiv f_{\rm{Fe}} = f_{\rm{Si}}$). Similar results were found by \citet{Guennou2015}. Hence, we have included the Si line in the $\rm{DEM}^{\rm{P}}$ calculations.
    Lastly, we used the new inferred $\rm{DEM}^{\rm{P}}_{Fe+Si}$ to simulate the intensity of the S \textsc{x} 264.233 \AA\ line and calculate the relative FIP bias
     \begin{equation}
        \frac{I^{\rm sim}_{\rm{S}}}{I^{\rm obs}_{\rm{S}}} = \frac{f_{\rm{(Fe+Si)}}}{f_{\rm{S}}}.
     \end{equation}

    \subsection{FIP bias measurement using linear combinations of intensities}
    There are other methods to calculate the FIP bias that do not rely on the derivation of the DEM.
    The oldest and most widely used method is the two-line-ratio method. This method assumes that when the contribution functions of one low-FIP line and one high-FIP line are similar, to some factor that can be approximated by max($C_{\rm LF}$)/ max($C_{\rm HF}$), the ratio $f_{\rm LF}/f_{\rm HF}$ becomes almost independent of the DEM. Then, the FIP bias is simply the ratio of intensities multiplied by a constant factor. Although the implementation of this method is straightforward, it is quite rare to identify such pairs of lines whose contribution functions match to a sufficient degree.

    The linear combinations ratio (LCR) method of  \citet{ZambranaPrado2019} is considered an extension of the two-line-ratio method and overcomes this problem by using a linear combination of spectral lines.
    The LCR method introduces new radiance-like and contribution-function-like quantities $\mathcal{I}$ and $\mathcal{C}$, respectively. These are the weighted linear combinations of the intensities and contribution functions, for the selected lines (see Equations 13, 14, 17 and 18 in \citealt{ZambranaPrado2019}). 
    
    By using a linear combination of line intensities and contribution functions with the appropriate coefficients, it is possible to refine the corresponding contribution functions for the low-FIP and the high-FIP elements, so that these match more closely with each other. Then, the relative FIP bias can be calculated from
    \begin{equation}
          \frac{f_{\rm{LF}}}{f_{\rm{HF}}} = \frac{\mathcal{I}_{\rm{LF}}}{\mathcal{I}_{\rm{HF}}} \left(\frac{\int \mathcal{C}_{\rm{LF}}\ \rm{DEM}\ dT_e} {\int \mathcal{C}_{\rm{HF}}\ \rm{DEM}\ dT_e} \right)^{-1},
    \end{equation}
    where the coefficients of the linear combination are the same for the $\mathcal{I}$ and $\mathcal{C}$.
    For a set of low-FIP and high-FIP lines the LCR methods optimizes their linear combination coefficients by trying to bring the ratio $\int \mathcal{C}_{\rm{LF}}\ \rm{DEM}\ dT_e/\int \mathcal{C}_{\rm{HF}}\ \rm{DEM}\ dT_e$ as close to one as possible for any DEM. This requires the use of ``reference" DEMs, for which the CHIANTI DEMs for quiet Sun, coronal hole, and active region are used \citep[for more details see][]{ZambranaPrado2019}.

    We found the optimal set of lines, for the implementation of the LCR method in our study, to be \{Fe \textsc{xi}, Si \textsc{x}, Fe \textsc{xii}, Fe \textsc{xiii}\}.
    A detailed description of the criteria for this selection is presented in Appendix.~\ref{app:LCR_lines}. 
    
\section{Results} \label{sec:results}

    \subsection{Longitudinal Cuts} \label{subsec:long_cuts}
    
    Our goal is to investigate the change in the FIP-bias values as we go from the coronal hole to the quiet Sun. 
    To do so, we took longitudinal cuts across the coronal hole and studied the FIP bias as function of distance from the coronal hole boundary.  
    In an ideal scenario we would have a set of continuous FIP-bias values along the cut. However, the FIP-bias values inside the coronal hole are not reliable. 
    This is because the intensities of some lines are close to the noise level inside the coronal hole. Additionally, previous studies \citep{Wendeln2018} have indicated that a significant portion of the line emission inside the coronal holes is due to light scattered from the surrounding quiet Sun. For these reasons, we excluded any coronal hole values from our analysis and focused on the behavior of the FIP bias outside the coronal hole. 
    The FIP-bias values, in this step of our analysis, were derived with the DEM method discussed in Section~\ref{subsec:fip_bias_dem}. 
    
\begin{figure*}[ht]
        \centering
        \includegraphics[scale=0.38]{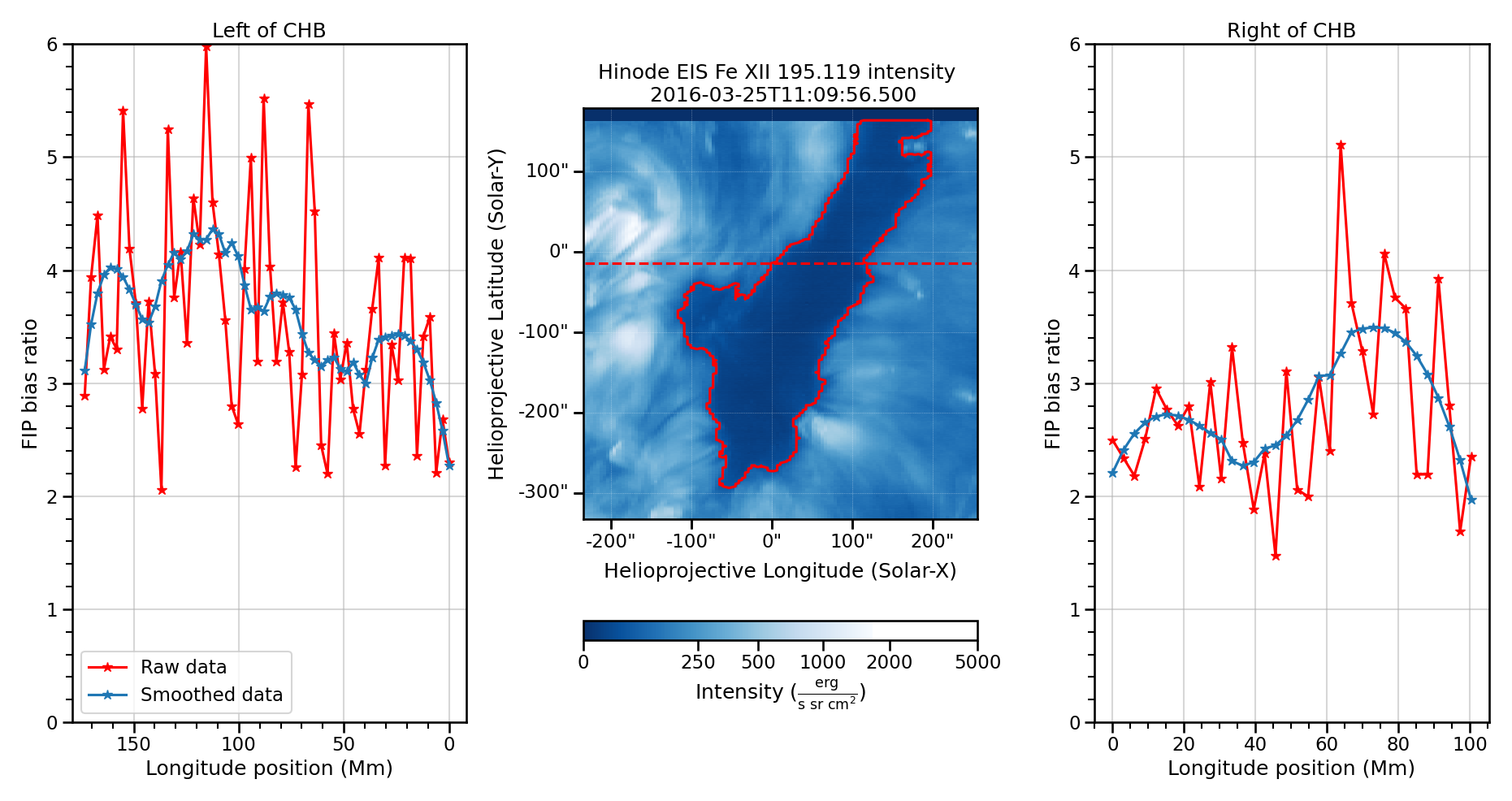} 
        \caption{FIP bias ratio along a longitudinal cut across the coronal hole. \textit{Left:} FIP bias data as a function of position along the longitudinal cut for the left side of the coronal hole boundary (CHB). \textit{Center:} The intensity of the Fe \textsc{xii} line including the calculated coronal hole boundary (red line). The red dashed line represents the longitudinal cut shown in the side panels. \textit{Right:} FIP bias data as a function of position along the longitudinal cut for the right side of the coronal hole.}
        \label{fig:long_cut}
    \end{figure*}

    To identify the coronal hole boundaries, we applied the triangle thresholding technique \citep{Zack1977} using the intensity of the Fe \textsc{xii} line for all pixels in the EIS FOV. 
    The triangle algorithm is a geometric method that is based on the histogram of the intensity (i.e., the $x$ axis is intensity and the $y$ axis the number of pixels where such intensity values occur). 
    Between the intensity value with the maximum number of occurrences ($b_{\rm max}$) and the intensity value with the minimum number of occurrences ($b_{\rm min}$), a line is drawn.
    Then, we calculate the perpendicular distance $d$ between this line and the histogram. 
    The intensity value $b_0$ at which the distance $d$ is maximal, is taken as the threshold value for determining the coronal hole boundary. We found that this method is more effective than other histogram-based thresholding techniques when there is only a weak peak in the histogram, as is often the case with coronal hole images in the EUV. We implemented this method using the Python package scikit-image \citep{VanDerWalt2014}. 
    An example of the derived coronal hole boundaries can be seen in the middle panel of Fig.~\ref{fig:long_cut}. 
  
    After identifying the coronal hole boundaries, we separated the FIP bias of the longitudinal cuts into two sections, one starting at the left edge of the coronal hole and going towards negative longitudes and one starting from the right edge and going towards positive longitudes. For each section of the cut, the longitudinal value of zero (on the $x$ axis) corresponds to the coronal hole boundary. 
    As can be seen in Fig.~\ref{fig:long_cut}, the unsmoothed FIP bias data along the two sections of the longitudinal cut display significant variability in their values. In order to investigate the overall trend of the FIP-bias values as a function of distance from the coronal hole boundary we applied a smoothing filter to the data. We used the scipy \citep{Virtanen2020} implementation of the Savitzky-Golay filter \citep{SavGol1964}. Before applying the smoothing, we took the Fourier transform of the FIP-bias values and identified the prominent spatial frequencies in the data. Based on that and the attenuation of the Savitzky-Golay filter as a function of the selected window size, we selected the filter window to be 20 pixels wide (with a polynomial of order 3).  
    The computed smoothed data, for a particular longitudinal cut can be seen in the two outer panels of Fig.~\ref{fig:long_cut}. For the rest of our analysis we use the smoothed values of the FIP bias.

    Figure~\ref{fig:long_cut} shows that to the left of the coronal hole, the FIP-bias values initially display quiet Sun values. Then, as we cross the loop system of the active region on the left of our FOV 
    (at approximately $140-100$ Mm) we see a significant increase. After the active region, the values drop back to quiet Sun values 
    (approximately at $100-50$ Mm), and as we approach the left edge of the coronal hole we see another decrease 
    ($50-0$ Mm). 
    This last portion is the focus of our study as it indicates the transition area for FIP-bias values from quiet Sun to coronal hole. We interpret this transition in the FIP-bias values as being related to a transition in the magnetic field topology from closed loops (quiet Sun) to open magnetic field (coronal hole). A similar behavior is observed in the longitudinal cut to the right of the coronal hole. There, we observe an increase in the FIP-bias values as we move away from the coronal hole boundary.  
    We note, that the conversion of the helioprojective coordinates (arcseconds) to physical distances on the solar surface (Mm) depends both on latitude and longitude. As our FOV is close to disk center and has a fairly symmetric extend in latitude and longitude we assumed a constant conversion factor for all the pixels inside our FOV. This introduces a negligible error, which peaks $\approx$ 4$\%$ at the corners of the FOV.

    \subsection{Average FIP bias Ratio}

\begin{figure*}[ht]
        \centering
        \includegraphics[scale=0.39]{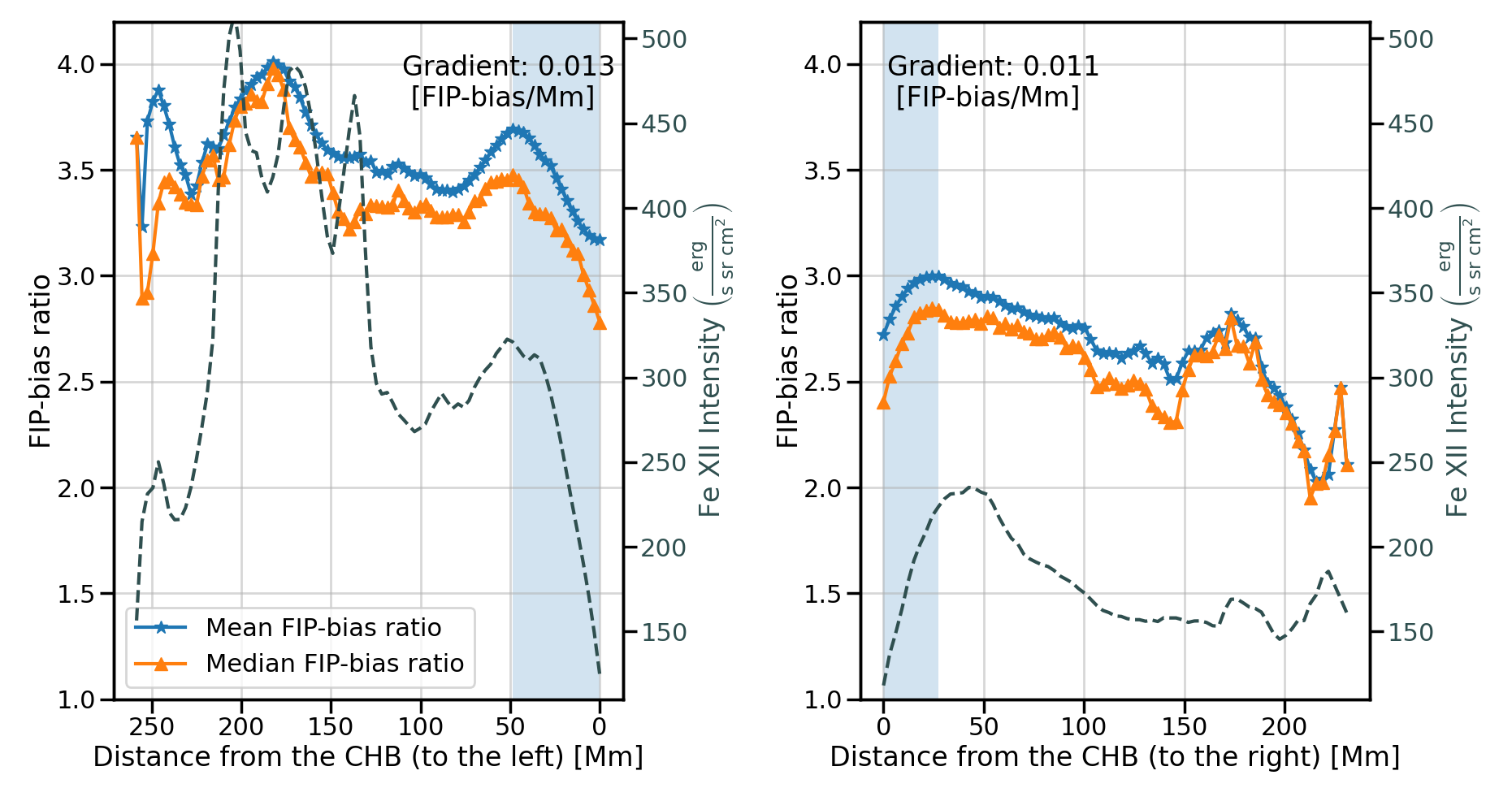}
        \caption{Average FIP bias as a function of distance (in longitude) from the coronal hole boundary. The blue shaded regions indicated where the interchange reconnection is inferred to occur. The FIP bias values are derived using the DEM method. The dashed line indicates the average intensity of the Fe \textsc{xii} line.}
        \label{fig:Average_FIP}
    \end{figure*}

    Continuing the DEM-method analysis, the behavior of the FIP bias close to the coronal hole for individual longitudinal cuts has a significant variability. This, stems mostly from the existence of different features along the cut. For example, if there is a small bright loop system very close to the coronal hole, as is the case at approximately $X = 60''$ and $Y = -220''$, large FIP-bias values adjacent to the coronal hole will be observed. 
    To address this variability, we focused our attention on identifying the overall trend of the FIP-bias values as we approach the coronal hole boundary, from each side. 
    We began by selecting from the smoothed FIP-bias data all longitudinal cuts that intersected the coronal hole within our FOV.
    Each cut was 1 pixel high in latitude.
    These longitudinal cuts were then divided into two parts---one corresponding to the left side and the other to the right side of the coronal hole, as is shown in Fig.~\ref{fig:long_cut}.
    Next, we aligned both portions for all longitudinal cuts with the corresponding left or right edge of the coronal hole. 
    We then averaged the FIP-bias values at each position away from the coronal hole, for both sides.
    Due to the morphology of the coronal hole, not all left and right portions of the longitudinal cuts have the same length. This means the further away we go from the coronal hole the fewer the number of longitudinal cuts that are used in the averaging. 
    The results of the average FIP bias, showing the mean and median values, as a function of distance for both sides of the coronal hole are presented in Fig.~\ref{fig:Average_FIP}.
    
    The average FIP bias displays a trend similar to what was seen for the individual longitudinal cut in Fig.~\ref{fig:long_cut}.
    This change in the FIP-bias values next to the coronal hole boundary is indicative of the area where we expect interchange reconnection to take place. In order to constrain the size of this area, we quantified the longitudinal extent of this transition using the averaged FIP-bias values. This was done taking the first derivative of the average FIP bias and identifying the position where it changes sign. We found that the width of this area is 50 Mm for the left side of the coronal hole and 30 Mm for the right side. These areas are indicated by the shaded portions in Fig.~\ref{fig:Average_FIP}.
    Furthermore, we examined how fast the average FIP-bias values change in these regions by applying a linear fit over that interval and deriving the slope. For both sides of the coronal hole the gradient of the linear fit was similar, with 0.013 FIP-bias/Mm for the left side and 0.011 FIP-bias/Mm for the right side. We also investigated if the variation in the FIP bias, as a function of distance from the coronal hole boundary, is related to the variation in the EUV intensity as observed from EIS. We expected that the EIS lines closer to the formation temperature of S \textsc{X} would mirror to a significant degree the variation of the FIP bias. Indeed, this behavior can be see in Fig.~\ref{fig:Average_FIP}, which shows the average intensity of Fe \textsc{XII}. This result indicates the potential use of such lines as a proxy for the variation of the FIP bias.
    
\begin{figure*}[ht]
        \centering
        \includegraphics[scale=0.39]{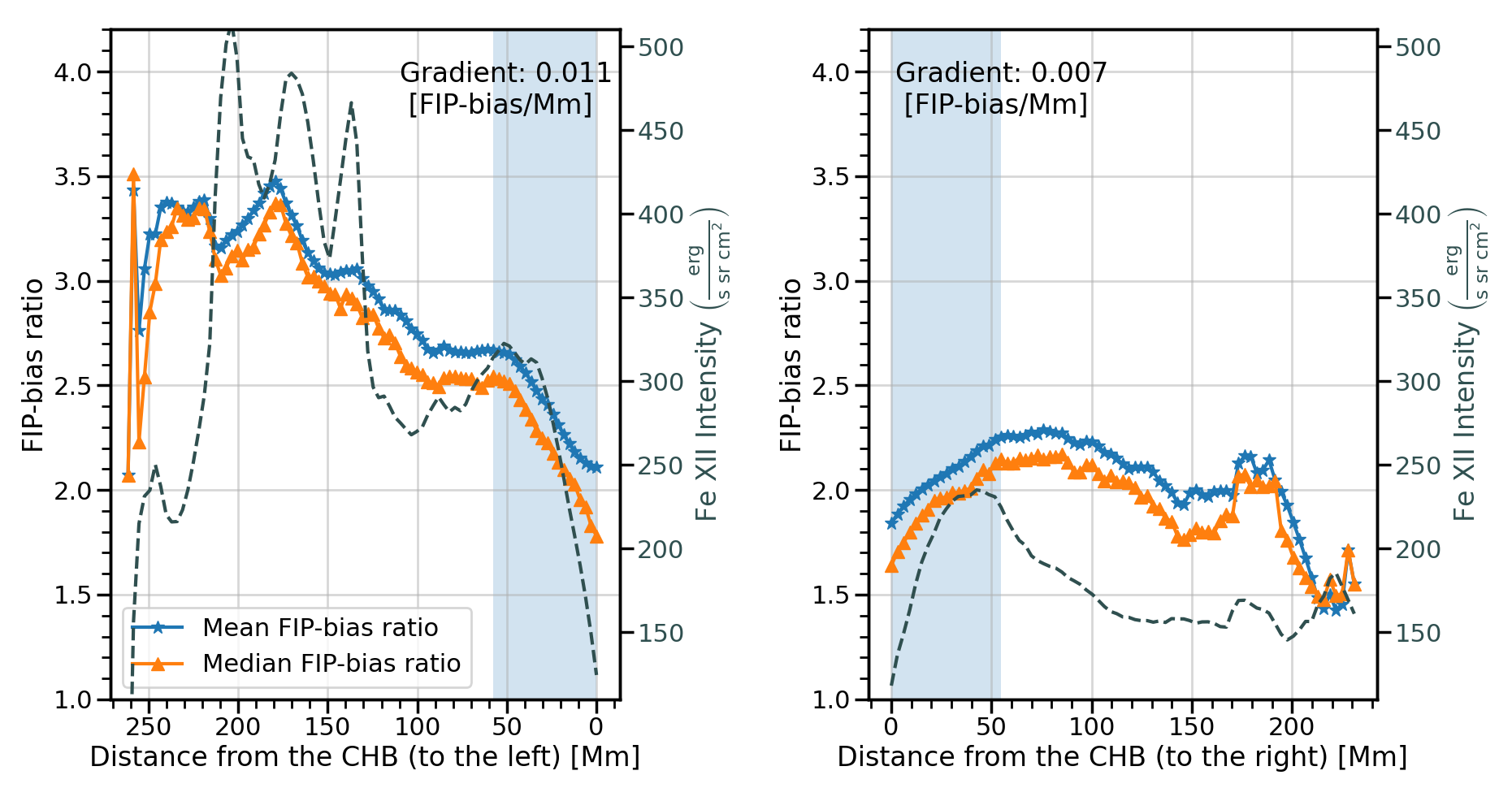}
        \caption{Same as Fig.~\ref{fig:Average_FIP} but using the LCR method.}
        \label{fig:Average_FIP_LCR}
    \end{figure*}

    Then, we repeated the above analysis using the LCR method and the lines: Fe \textsc{xi}, Fe \textsc{xi}, Si \textsc{x}, Fe \textsc{xii}, Fe \textsc{xii}. The results are presented in Fig.~\ref{fig:Average_FIP_LCR}. 
    The LCR method gives coronal hole boundary widths of 58 and 55 Mm, for the left and the right side of the coronal hole, respectively. 
    These areas are shown by the shaded portions in Fig.~\ref{fig:Average_FIP_LCR}.
    The gradients from the linear fit in these areas are 0.011 FIP-bias/Mm for the left side and 0.007 FIP-bias/Mm for the right side. 
    The LCR method produces smaller FIP-bias values than the DEM method and slightly smaller gradients. But both methods show a clear increase in the FIP-bias values as we move away from the coronal hole boundary.  
    Possible reasons for the discrepancy in the FIP-bias values derived from the LCR method include the specific combination of lines selected (see Appendix.~\ref{app:LCR_lines}) and the use of a reference DEM, which may not accurately represent the real conditions. 
    A detailed analysis of why the FIP-bias magnitudes systematically differ between the DEM and LCR methods is beyond the scope of this paper.
    Nevertheless, despite these differences, the width of the boundary region and the slope of the FIP bias agree well between the two methods, providing confidence in the robustness of our results.
    
    \subsection{Diffusion coefficient} \label{subsec:diff_coeff}
    We can use the properties of the average FIP bias in the boundary region around the coronal hole to derive an estimate for the diffusion coefficient of the open flux diffusion model. 
    The coronal hole boundary width in the FIP-bias values (the shaded areas in Figs.~\ref{fig:Average_FIP} and \ref{fig:Average_FIP_LCR}) can be interpreted as an upper limit for the distance that an open field line jumps during interchange reconnection, $\delta h_{\rm max}$. 
    The FIP bias builds up over time in closed loops, at a rate of about 1/day, and is a proxy for how long a magnetic field line remained closed. Consequently, the magnitude of the FIP bias in the boundary region around the coronal hole is proportional to the time since reconnection between an open and a closed field line, which is in turn proportional to the characteristic reconnection time $\delta t$. 
    Hence, we can consider the mean FIP bias in the coronal hole boundary region, $\left<{\rm FIP bias}\right>$, to be an estimate of the average reconnection time, $\delta t_{\rm avg} = [\left<{\rm FIP bias}\right> - 1]$~day.
    
    Using these simplifications for $\delta h$ and $\delta t$, we can compute the diffusion coefficient for our results with the DEM method
    \begin{equation*}
        \begin{split}
            \kappa_{\mathrm{left}} & = 6.7  \times 10^{13}\ \mathrm{cm}^2 \ \mathrm{s}^{-1}, \\
            \kappa_{\mathrm{right}} & =  3.9 \times 10^{13}\ \mathrm{cm}^2 \ \mathrm{s}^{-1}, \\
            \kappa_{\rm DEM} & = 5.3 \times 10^{13}\ \mathrm{cm}^2\ \mathrm{s}^{-1}, \\
        \end{split}
    \end{equation*}
    and with the LCR method
    \begin{equation*}
        \begin{split}
            \kappa_{\mathrm{left}} & =  1.6 \times 10^{14}\ \mathrm{cm}^2 \ \mathrm{s}^{-1}, \\ 
            \kappa_{\mathrm{right}} & =  1.9 \times 10^{14}\ \mathrm{cm}^2 \ \mathrm{s}^{-1}, \\
            \kappa_{\rm LCR} & = 1.75  \times 10^{14}\ \mathrm{cm}^2\ \mathrm{s}^{-1}. \\
        \end{split}
    \end{equation*}
    Combining the results from both methods, we find $\kappa = 1.3 \pm 0.6 \times 10^{14}\ \mathrm{cm}^2\ \mathrm{s}^{-1}$.

\section{Discussion} \label{sec:discussion}

    \subsection{Comparison of the diffusion coefficient}

    \citet{Fisk2001}, with their theoretical model, calculated $\kappa = 1.6 \times 10^{15}\ \mathrm{cm}^{2}\ \mathrm{s}^{-1}$ outside coronal holes, for low latitudes during both solar minimum and maximum. Inside polar coronal holes, for solar minimum, they found $\kappa = 3.5 \times 10^{13}\ \mathrm{cm}^{2}\ \mathrm{s}^{-1}$. 
    For their calculations, they treated $\delta h$ as being a significant fraction of the loop height. Based on observations by \citet{Feldman2000}, \citet{Fisk2001} considered typical loop heights of $20-400$ Mm. Building on that, they selected an average loop height of 200 Mm. 
    By extrapolating solar wind measurements at 1 AU \citep{Fisk1999}, they found the characteristic reconnection time to be 38 hours for low latitudes and 36 hours for polar coronal holes. These values were comparable to the observed timescale over which concentrations of open flux are enhanced or disappear \citep{Schrijver1998}. 
    
    Our results for the diffusion coefficient, at the coronal hole boundary region, lie between the \citet{Fisk2001} values outside and inside coronal holes. This could be an indication that coronal hole boundary regions represent an intermediate configuration for the open field diffusion between coronal holes and quiet Sun. Such an intermediate configuration is also seen in the FIP bias values. To study the nature of this intermediate state we examine more closely the factors that contribute to the diffusion coefficient.
    The average reconnection time that we derived from both methods, DEM and LCR, is $\approx$ 36 hours.
    This implies that the main reason for the discrepancy is the value of $\delta h$.
    The loop heights in \citet{Fisk2001} were based on the observations by \citet{Feldman2000}, who showed that in the upper atmosphere of quiet Sun regions there are large and hot loops ($\approx 1.4 \times 10^6$ K) that create a canopy over smaller and cooler loops ($\approx 1.0 \times 10^6$ K). 
    Additionally, \citet{Feldman2000} found that the smaller and cooler loops show a correlation with the supergranule network, which was not the case for the larger and hotter loops.
    Following the assumption of \citet{Fisk2001}, we interpret our results of $\delta h =  30-60$~Mm as a proxy for the heights of the loops involved in interchange reconnection. 
    Under this scenario, our results at the coronal hole boundary region imply that interchange reconnection takes place predominantly with the smaller loops. 
    However, that interchange reconnection would preferentially occur only with loops of certain heights seems unphysical.
    
    A different interpretation would be that our results are affected by an observational bias that restricts our findings only to reconnection close to the coronal hole boundary, due to our limited FOV. 
    Thus, we are able to distinguish reconnection with the smaller loops, which reconnect to create new loops within $30-60$ Mm of the coronal hole; but any reconnection with much larger loops, $> 200$~Mm, would mean that the footpoints of the newly formed loops lie at the edge or outside our FOV. 
    Some initial supporting evidence for this interpretation could be the decrease in the average FIP-bias values that we observe close to the edge of our FOV on the right side of the coronal hole (see Fig.~\ref{fig:Average_FIP} right panel). 
    This decrease could be an indication for a small concentration of open flux 
    formed by reconnection with larger loops. 
    We note, though, that due to the tilt of the coronal hole as we approach the edge of our FOV, the average FIP bias is calculated by a smaller number of datapoints. This means that trends observed near the edge of our FOV are related to a small region and may not representative of the overall behavior. 
    We are not able to distinguish if there is a similar behavior for the left side of the coronal hole due to the presence of the active region.
    Additionally, as we move farther from the coronal hole, we sample a larger area, which reduces the concentration of the open flux, making it more difficult to distinguish from the surrounding quiet Sun. 
    We will study this interpretation further in future work using observations with a larger FOV.

    \subsection{Connection with the Open Flux Problem}

    The interplanetary magnetic field (IMF) is the extension of the solar magnetic field into the heliosphere and can be measured in situ by spacecraft. The total open flux of the Sun can be estimated from these spacecraft measurements \citep{Owens2008}. 
    A complimentary estimate of the total open flux can be derived from remote sensing measurements of the photospheric magnetic field combined with coronal magnetic field models \citep{Linker2017a}. 
    However, there is a discrepancy in the magnitude of open flux derived from the two methods, with the magnetograms/models method being a factor of $2-4$ smaller than the in situ estimates. This is know as the open flux problem. 
    There has been considerable effort to reconcile the results between the magnetograms/models method and the in situ data \citep[and references therein]{Arge2024}. 
    Observations outside the ecliptic, from Ulysses, showed that there is a latitudinal invariance in the magnitude of the radial interplanetary magnetic field \citep{Smith2003,Lockwood2004}. This has also been confirmed by other broader studies \citep{Owens2008}. Thus, the in situ results represent direct measurements of the open flux and the main considerations for the discrepancy have focused on the magnetograms/models results. These considerations have been two fold. One is that there is an underestimation of the open flux in the magnetogram maps from different observatories \citep{Wang1995,Riley2014}. The other is that we need to more accurately determine the size of open flux areas on the Sun, which are typically determined from dark regions in EUV observations. 
    Related to this, \citet{Linker2017a} showed that regions of open magnetic field need to exceed the coronal hole areas, by $\sim$70\%, in order for any magnetogram/model configuration to match the in situ open flux calculations. 
    However, \citet{Linker2021a} also found that the uncertainty in the EUV-estimated coronal hole areas can account for only a small portion of the missing flux and is not sufficient to explain the factor of $2-4$ difference compared to the in situ estimates. 

    Another factor that can contribute to the total amount of unaccounted open flux from the magnetograms/models method is if a portion of the open flux is rooted in regions that do not appear as dark as coronal holes in EUV emission. 
    The diffusion of open flux due to reconnection at the boundary regions of coronal holes, is one such potential source of open flux. Our results indicate that we have opening and closing of magnetic field lines in an area $\sim 30 - 60$ Mm wide next to the coronal hole. 
    Consequently, there is an amount of diffused open flux in this area (i.e., in the coronal hole boundary width), which appears brighter than the coronal hole in EUV observations. 
    
    \begin{figure}[ht]
        \centering
        \includegraphics[scale=0.39]{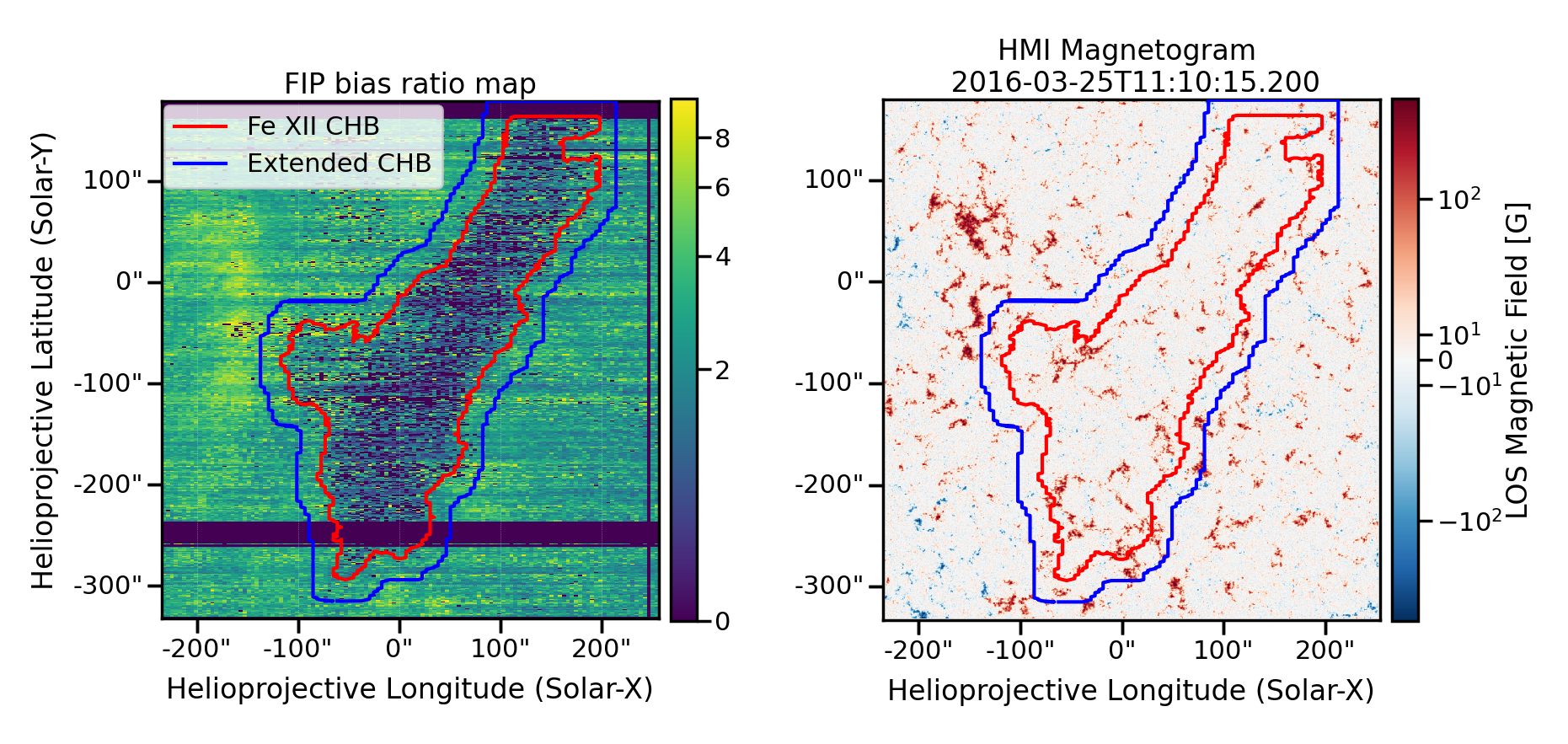}
        \caption{The coronal hole boundary and the extended coronal hole boundary. \textit{Left:} FIP bias ratio map with contours showing the coronal hole boundary from Fe \textsc{xii} (in red, see also Fig.~\ref{fig:long_cut}) and the extended coronal hole boundary in blue, for a coronal hole boundary width of 30 Mm. Portions of our FOV, where the data are corrupted, have been set to zero and are visible as dark strips. \textit{Right:} LOS HMI magnetogram aligned with the EIS FOV and with the same contours.}
        \label{fig:FIP_map_CHB}
    \end{figure}
   
    We can derive a first order estimate for the amount of this diffused open flux. First, we re-project the coronal hole boundary (left panel in Fig.~\ref{fig:FIP_map_CHB}) to the closest-in-time LOS magnetogram (right panel in Fig.~\ref{fig:FIP_map_CHB}). The magnetogram was retrieved from the Helioseismic and Magnetic Imager \citep[HMI;][]{Scherrer2012}. We calculate the total area of the coronal hole, $\rm{A}$, and use the magnetic field, $\rm{B}$, inside this area. The open magnetic flux of this coronal hole is then given by $\Phi = \rm{B} \times \rm{A}$. 
    Next, we find the area of the coronal coronal boundary region. 
    Based on our analysis of the FIP-bias values we selected widths for the coronal hole boundary region of 30 and 60 Mm. 
    We extended the coronal hole boundary by implementing a binary dilation on the coronal hole boundary mask. For the binary dilation we used the scipy package and a rectangular structuring element that followed the resolution of our observations.  
    The original coronal hole boundary and the extended coronal hole boundary, for a width of 30~Mm, are displayed in Fig.~\ref{fig:FIP_map_CHB}. 
    The area between these two boundaries contains some residual diffused open flux, based on the FIP bias results. In order to simulate the transition from fully open to fully closed flux, we consider that the open flux $\Phi$ of the coronal hole, drops linearly to 0 from the edge of the coronal hole boundary to the edge of the extended coronal hole boundary. 
    Under this assumption, for coronal hole boundary widths of $\sim 30-60$~Mm we find 
    \begin{equation}
        \frac{\Phi_{\rm{extended-CH}}}{\Phi_{\rm{CH}}} \approx 1.37 - 1.71
        \vspace{0.1cm}
    \end{equation}
    indicating a $37 - 71\%$ increase in the amount of open flux. 
    These results, although a first approximation, indicate that diffused open flux next to the coronal hole boundary can have a measurable contribution. Though the estimates for the amount of this flux do not appear to be sufficient to reconcile the difference with the in situ calculations, they help to reduce the gap and may point to a resolution for the open flux problem. 

    Additionally, the magnitude of open flux in the coronal hole boundary region can be even larger in cases where the coronal hole is located close to an active region. 
    At such configuration the footpoints of the new open field lines after interchange reconnection will end up in the vicinity of the active region.
    As active regions have stronger magnetic field, they would provide a much larger amount of diffused open flux and this could further reduce the discrepancy between the magnetogram/models and in situ results. Unfortunately, our current FIP bias measurements are not sufficient to resolve any observational signature of this topology for the coronal hole we studied.
    However, the recent work of \citet{Arge2024} provides an encouraging indication for the validity of this interpretation. They showed that there is a significant amount of open flux in the boundary of coronal holes, by examining how the magnetic field lines are traced inside a model of the coronal magnetic field and how the total unsigned open flux is estimated. During solar maximum, the previously unaccounted open flux cells in the boundary of coronal holes lie close to active regions. This results in a larger amount of open flux from these locations, due to the strong magnetic fields of active regions. Their estimation for the size of these cells is $\sim$ 24~Mm, which is comparable to our results. Lastly, we examined the magnetic field data for any distinct differences between the coronal hole area and the coronal hole boundary regions with widths of 30 and 60 Mm. The photospheric magnetic field displayed similar properties within the coronal hole region and the coronal hole boundary region with a width of 30 Mm. Some differences in magnetic field strength were seen between the coronal hole and the boundary region with a width of 60 Mm but we attribute this to the larger boundary region encompassing part of the loop system to the left of the coronal hole. In a future work we will perform the more detailed analysis required to draw definitive conclusions about the magnetic field topology and its relationship to interchange reconnection.

\section{Conclusions} \label{sec:conclusions}

In this work, we have computed the FIP bias at the boundaries of an equatorial coronal hole, using the DEM and LCR methods. We have examined the average behavior of the FIP bias as a function of distance from the coronal hole boundary. This was done by identifying the coronal hole area and taking longitudinal cuts at all latitudes, within our FOV, that intersect with the coronal hole. 

Our results showed that the boundary region, where the FIP bias goes from coronal hole values to quiet Sun values, has a width that ranges from $\sim 30 - 60$ Mm. The inferred coronal hole boundary width has some uncertainty, depending on the method that we used to calculate the FIP bias; but overall it is comparable to the $\approx$ 30~Mm size of supergranules. 
We found that the coronal hole boundary width was slightly larger for the left side of the coronal hole compare to the right, for both methods, but this variation was not significant.  

Furthermore, we converted the properties of the average FIP bias in the coronal hole boundary region to an estimate of the diffusion coefficient, using the open flux diffusion model.
Our results were an order of magnitude smaller than the theoretical estimate of \citet{Fisk2001}. 
To examine more closely this discrepancy, we followed the assumption of \citet{Fisk2001}, that the distance an open field footpoint jumps after reconnection is proportional to the loop height. Under this assumption, we attributed this observed discrepancy to the size of the loops that take place in the interchange reconnection. \citet{Fisk2001} assumed an average loop height of 200 Mm, whereas our results indicated that smaller loops of $\sim 30 - 60$ Mm heights facilitate the reconnection at the coronal hole boundary region. 
It is possible, in our selected observations, that larger loops of heights $> 200$ Mm also contribute to the reconnection with the coronal hole magnetic field. That could potentially account for some of the discrepancy between our findings and those of \citet{Fisk2001}, 
but any such signature of reconnection via such loops is at the edges or outside our FOV. 
The validity of this interpretation will be tested in the future with the analysis of additional events using larger FOVs.

Our results also provide a new insight into the open flux problem.
Based on the boundary region inferred from the FIP bias, there is an area next 
to coronal holes with diffused open flux. This flux, rooted in regions that are brighter in EUV emission than the coronal holes, was previously not considered in the estimates of the total open flux of the Sun. With our first order approximation for the amount of this diffused open flux, we were able to show that it can have a significant contribution, $\sim 37 - 71\%$, to the amount of the unaccounted open flux. This interpretation is supported by the recent work of \citet{Arge2024}, who improved the tracking of open flux in coronal magnetic field models. Additional unaccounted open flux could as well come from the $> 200$~Mm loops discussed above. 

Our planned future studies will entail the analysis of more events using a larger FOV. This will help to test the robustness of our results, examine in more detail any significant variation with latitude, and explore for the role of footpoint jumps of $> 200$~Mm. Furthermore, we plan to examine the mapping of in situ solar wind and how the solar wind properties relate to the FIP bias measurements. 
Such studies will be crucial in refining our understanding of the solar open flux distribution, offering insights into the origins of the unaccounted open flux and its role in shaping the formation of the slow solar wind.   

\vspace{1cm} %fix for Arxiv only
% \begin{acknowledgments} 
    \textit{Acknowledgments:} A. Koukras, D. W. Savin, and M. Hahn are supported, in part, by the NASA Heliophysics Living With a Star (LWS) program grant 80NSSC20K0183.
    CHIANTI is a collaborative project involving George Mason University, the University of Michigan (USA), University of Cambridge (UK) and NASA Goddard Space Flight Center (USA).
    The authors thank Natalia Zambrana Prado, Stefan Hofmeister, and Peter Young for the interesting discussions and their comments.  
% \end{acknowledgments}

\software{Suny \citep{Mumford2015}, Astropy \citep{Astropy2018}, Scipy \citep{Virtanen2020}, scikit-image \citep{VanDerWalt2014}, Matplotlib \citep{Hunter2007}, EISPAC \citep{Weberg2023}}
\facilities{Hinode (EIS), SDO (HMI, AIA)}

\bibliography{references}{}
\bibliographystyle{aasjournal}

\appendix

\section{Line selection for LCR method} \label{app:LCR_lines}
As stated in \cite{ZambranaPrado2019}, the selection of the lines to be used in the LCR FIP bias calculation has to be done by hand. 
Although they provide a number criteria that the selected lines should satisfy, these are not sufficient to determine whether one set of lines should be preferred over another. In order to have a more quantifiable justification, we examined how well 
the linear combination contribution function of the low-FIP lines match with the contribution function of our high-FIP line. For this comparison, we focused mainly on the results for the quiet Sun reference DEM, as this is the most representative of the conditions in our FOV.

\begin{figure}[hb]
    \centering
    \includegraphics[scale=0.5]{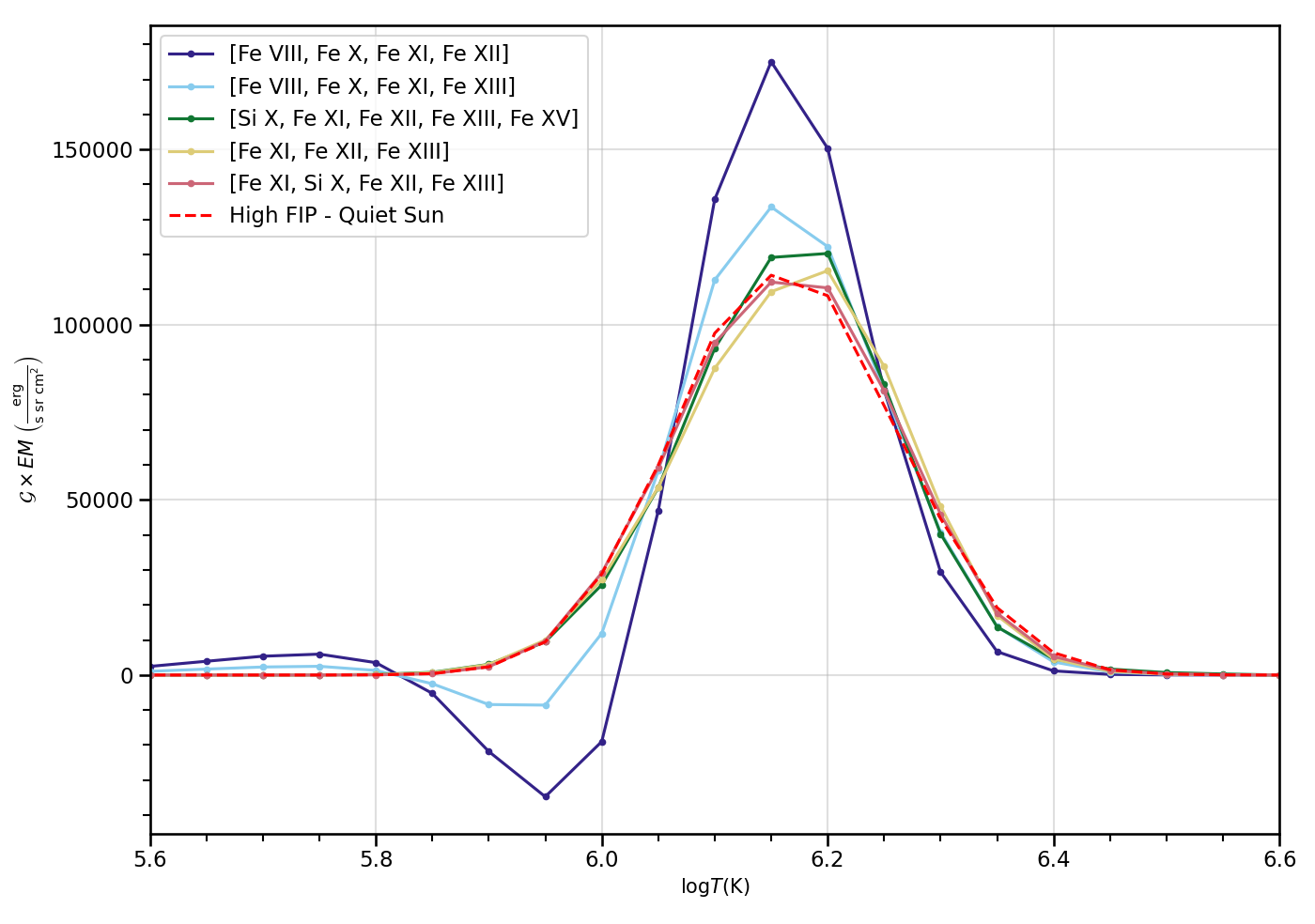}
    \caption{Comparison of the $\mathcal{C}_{\rm LF} \times EM$ product for different combinations of lines and our high-FIP element as a function of temperature $T$.
    Here, we focus on the quiet Sun reference DEM, but similar results were obtained for the active region and coronal hole reference DEMs.}
    \label{fig:LCR_best_comb}
\end{figure}

\begin{figure}[hb]
    \centering
    \includegraphics[width=\textwidth]{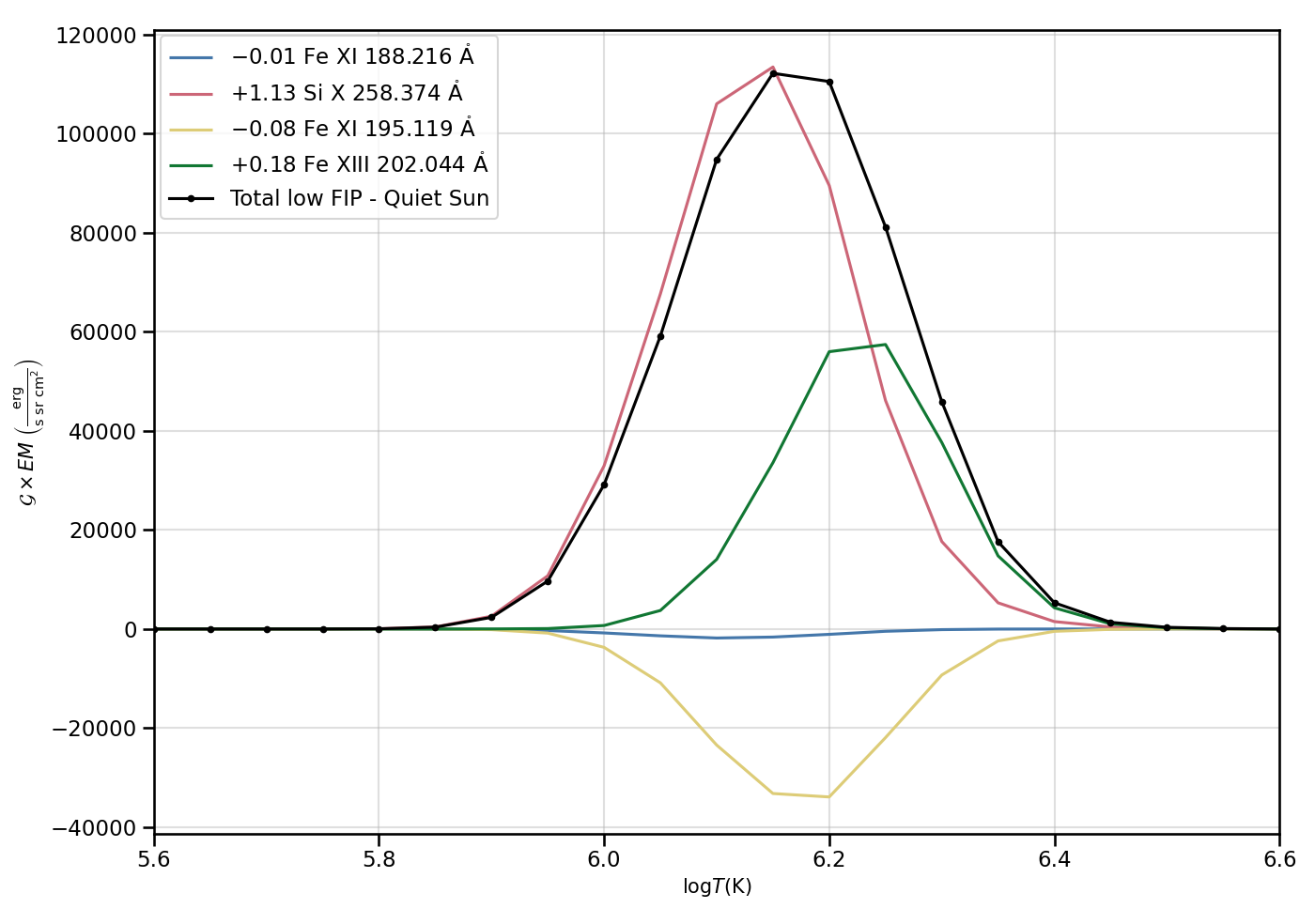}
    \caption{Contribution of each line to the $\mathcal{C}_{\rm LF} \times EM$ product as a function of temperature for the quiet Sun reference DEM, for the linear combination of [Fe \textsc{xi}, Si \textsc{x}, Fe \textsc{xii}, Fe \textsc{xiii}]. The total $\mathcal{C}_{\rm LF} \times EM$ product is shown with the black line. The LCR coefficient of each line is given in the legend.} 
    \label{fig:LCR_best_comb_coeff}
\end{figure}

We examined multiple combinations of lines. The 5 combinations with the best matches are displayed in Figure~\ref{fig:LCR_best_comb}. The combination [Fe \textsc{xi}, Si \textsc{x}, Fe \textsc{xii}, Fe \textsc{xiii}] produces the best match both for the quiet Sun reference DEM and for the coronal hole and active region DEMs. 
The contribution of the each line in the optimal set of lines are shown in Figure~\ref{fig:LCR_best_comb_coeff}.

\end{document}